\documentclass[aps,prl,reprint,groupedaddress]{revtex4-1}

\usepackage{graphicx}
\usepackage{bm}

\begin{document}


\title{Mode competition and anomalous cooling in a multimode phonon laser}


\author{Utku Kemiktarak$^{1,2}$, Mathieu Durand$^{2}$, Michael Metcalfe$^{1,2}$, and John Lawall$^{2}$}
\address{$^1$Joint Quantum Institute, University of Maryland, College Park, MD 20742, USA \\
$^2$National Institute of Standards and Technology, 100 Bureau Drive, Gaithersburg, MD 20899, USA}

\date{\today}

\begin{abstract}
We study mode competition in a multimode ``phonon laser'' comprised of an optical cavity
employing a highly reflective membrane as the output coupler.  Mechanical gain is provided by 
the intracavity radiation pressure, to which many mechanical modes are coupled.  We calculate the gain, and find that strong oscillation in one mode suppresses the gain in other modes.  For sufficiently strong oscillation, the gain of the other modes actually switches sign and becomes damping, a process we call ``anomalous cooling.'' We demonstrate that mode competition leads to single-mode operation and find excellent agreement with our theory, including anomalous cooling.
\end{abstract}


\maketitle

{\em Introduction. -- }
While the laser was invented more than five decades ago, its acoustic analog 
has only recently been realized.  Following the observation of phonon amplification
in microwave-pumped ruby\cite{Tucker_Ruby} in 1964, the possibility of ``phonon lasing'' was suggested.  Subsequent work in ruby
studied the emission spectrum of phonon generation\cite{ganapolskii_resonant_1974}
and multimode processes\cite{ganapolskii_resonant_1994}.  With the advent of optical pumping,
detailed studies of the coherence of phonon emission were enabled\cite{tilstra_coherence_2003},
culminating in a ruby ``saser''\cite{tilstra_optically_2007}.
Shortly thereafter, phonon lasing was realized in a harmonically bound Mg$^+$ ion driven by optical 
forces\cite{vahala_phonon_2009}.  Subsequently, it was recognized that a large class of optomechanical systems, in which optically furnished gain enables self-sustained mechanical oscillation, are properly 
called ``phonon lasers''\cite{khurgin_laser-rate-equation_2012}.  
These include beams\cite{khurgin_laser-rate-equation_2012,bagheri_dynamic_2011} and cantilevers\cite{metzger_self-induced_2008} coupled to an optical cavity, microtoroids\cite{kippenberg_cavity_2007,grudinin_phonon_2010}, and a cantilever deriving mechanical gain from optical bandgap excitation\cite{okamoto_vibration_2011}.  
Analogous electromechanical\cite{bargatin_nanomechanical_2003,Bennett2006} and purely mechanical
\cite{mahboob_phonon_2013} systems have also been discussed.
Various phenomena associated with lasers, such as stimulated emission\cite{vahala_phonon_2009}, oscillation threshold\cite{vahala_phonon_2009,khurgin_laser-rate-equation_2012,grudinin_phonon_2010,okamoto_vibration_2011,mahboob_phonon_2013}, gain narrowing\cite{mahboob_phonon_2013}, and injection locking\cite{knuenz_injection_2010}, have been demonstrated.


With few exceptions,
these investigations have involved a single mechanical mode.  
Multimode emission was observed in ruby\cite{ganapolskii_resonant_1974,ganapolskii_resonant_1994}, and two-mode oscillation was observed in a photothermally coupled optomechanical system\cite{metzger_self-induced_2008}.  Intermodal coupling in an electromechanical
system was exploited to realize a phonon laser without an optical pump\cite{mahboob_phonon_2013}. In the domain of conventional
lasers, an interesting and important feature arises when multimode operation is considered.  As shown by 
Lamb in 1964\cite{lamb_theory_1964}, a saturation phenomenon
occurs in which oscillation of one mode suppresses the gain of other modes.  This has the dramatic consequence  that, in the absence of inhomogeneous gain broadening, a laser oscillates in steady state on a single mode, 
even when the small-signal gain exceeds the losses for more than one mode\cite{Siegman}.  

In view of the multimode oscillation observed in the photothermal system\cite{metzger_self-induced_2008}, it is natural to ask 
whether a phonon laser employing pure radiation pressure coupling would exhibit the single-mode oscillation characteristic of a homogeneously broadened laser.
Here we study such a system, in which one cavity mirror is formed by a highly
reflective membrane supporting many mechanical modes.  We find that when the mechanical gain exceeds the losses
for more than one mode, the steady-state condition is nevertheless always that of a single oscillating mode. 
We calculate the gain for the case of two modes, and find that, just as in the conventional laser, a strongly
oscillating mode tends to ``steal'' gain from competing modes.
Sufficient oscillation amplitude, in fact, reverses the sign of the gain of more weakly oscillating modes, causing them to be optically damped.  
Experimentally, we are able to force
certain modes to oscillate, or to put the system in a regime in which the mode that ultimately oscillates
is unpredictable and depends on thermal fluctuations.  In addition, we verify our prediction
that as the oscillation strength of one mode is increased, the quenched modes are in fact cooled.

\begin{figure}
\includegraphics[width=3.38in,keepaspectratio]{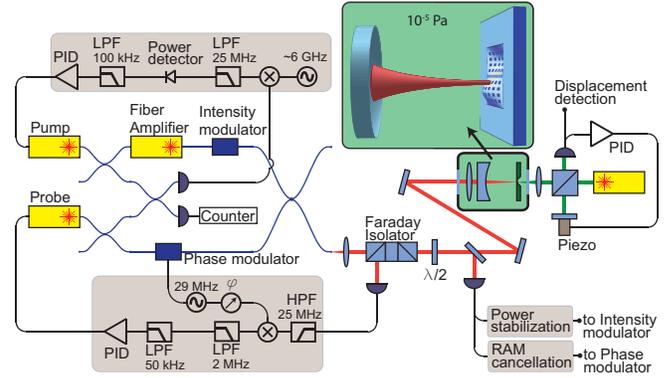}
\caption{A Fabry-Perot cavity employing a highly reflective sub-wavelength grating
in a silicon nitride membrane, operated in vacuum, forms the optomechanical system.  A ``probe'' laser is locked to a cavity
resonance, while a ``pump'' laser is blue-detuned relative to the
adjacent longitudinal cavity resonance and provides mechanical gain.  
A third laser is used for a Michelson interferometer to sense membrane displacements.}
\label{fig:fig1}
\end{figure}

{\em Experiment. -- }
As illustrated in Fig.~\ref{fig:fig1}, our optomechanical resonator is used as the output coupler of a Fabry-Perot cavity.  It is a square (side $a=1.25\,\mathrm{mm}$) silicon nitride membrane, patterned with an array of gratings, each 
$50\,\mu\mathrm{m}$ on a side.  Each grating has a period smaller than the wavelength ($1.56\,\mu\mathrm{m}$) of the light, and has near-unity reflectivity.  
The optomechanical properties of the device have been reported previously\cite{kemiktarak_mechanically_2012,Kemiktarak_NJP_2012}. Only one of the gratings is used at a time, but by choosing
a particular grating, one can optimize the radiation pressure coupling to a certain set of mechanical modes.
All of the results in this paper employ a grating centered at $(x_0,y_0)=(465(10)\,\mu\mathrm{m},400(10)\,\mu\mathrm{m})$ relative to a membrane corner. 
The reflectivity of this particular grating yields a cavity finesse $F\approx 1000$.  
The input coupler has a radius of curvature of $R=25$~mm. 
We use a cavity length $L_{cav}\approx R$ so that the waist $\omega_0$ of the optical mode is below $20\,\mu\mathrm{m}$; correspondingly, the cavity free spectral range is $\Delta\nu\approx6\,\mathrm{GHz}$.   

We introduce two lasers into the cavity: a probe laser and a pump laser, both with $\lambda=1.56\,\mu\mathrm{m}$,
as shown in Fig.~\ref{fig:fig1}.  The probe laser is locked to a cavity resonance; 
the pump is frequency-offset from the probe laser by $\Delta\nu+\delta\nu$, where $\delta\nu$ represents
an arbitrary detuning from the adjacent cavity mode.  
In order to transduce membrane displacements, we implement a Michelson 
interferometer targeting the grating used for the optical cavity.  The Michelson beam employs a third laser 
at $1.56\,\mu\mathrm{m}$ that is far off resonance from any cavity mode (cavity linewidth$\,\approx6\,\mathrm{MHz}$). 

{\em Radiation pressure-induced dynamics. -- }
The power circulating in the Fabry-Perot cavity is correlated to the membrane motion, optically modifying the dynamics.
The case of a single mechanical mode has been studied extensively\cite{kippenberg_cavity_2007}; the radiation pressure enables optically modified frequency shifts, cooling, and oscillation.
Here we generalize to multiple mechanical modes.
We express the membrane displacement $z(x,y,t)$ as a sum of products of normal modes $\phi_{mn}(x,y)$ with time-dependent factors $q_{mn}(t)$:
$z(x,y,t)=\sum_{m,n}q_{mn}(t)\phi_{mn}(x,y)$. 
For a uniform square membrane, the  modes  are given by 
$\phi_{mn}(x,y)=\sin\frac{m\pi x}{a}\sin\frac{n\pi y}{a}$,
and the effective mass $m_{eff}$ is equal to one fourth of the membrane 
mass\cite{Timoshenko}.
Each mode is driven by generalized force 
$F_{mn}(t)=\int\int f(x,y,t)\phi_{mn}(x,y)\,dx\,dy$\cite{Timoshenko},
where $f(x,y,t)$ is the radiation pressure force per unit area.

The amplitude $u(t)$ of the electric field circulating in a high-finesse Fabry-Perot cavity with a varying cavity length 
$L(t)=L_0+z(t)$ is governed by the differential equation
\begin{equation}
\dot{u}(t)+\left(\gamma - i\,(\delta\omega+\frac{4\pi z(t)}{\lambda} \Delta\nu)\right)\,u(t)=i\Delta\nu\,\sqrt{T_1\,P_{in}} 
\label{eqn: field evolution}
\end{equation}
where $\Delta\nu=c\,/(2L_0)$, $\gamma$ is the cavity field decay rate, related to the
finesse~$F$ by $\gamma=\pi\Delta\nu/F$, and $T_1$ is the transmission of the input coupler. 
The cavity is driven by laser light of frequency $\nu_L$, wavelength $\lambda$, and power $P_{in}$, detuned from a resonance frequency $\nu_0$ by $\delta\omega=2\pi\delta\nu=2\pi(\nu_L-\nu_0)$.  
The intensity distribution is Gaussian, with spot size $\omega_0$, centered at $(x_0,y_0)$.  

For a sinusoidal membrane oscillation $q_{mn}(t)\equiv z_{mn}\sin 2\pi\nu_{mn} t$,  the solution to Eq.~(\ref{eqn: field evolution}) contains a spectrum of sidebands separated by $\nu_m$.
The dimensionless quantity $\chi_{mn}\equiv 2\,\frac{\Delta\nu}{\nu_{mn}}\frac{z_{mn}}{\lambda}\phi_{mn}(x_0,y_0)$ appears as a natural expansion parameter, and, for $\chi_{mn}>1$, corresponds roughly to the number of sidebands with significant amplitude.
The radiation pressure $F^{RP}=2|u(t)|^2/c$ associated with the circulating optical power oscillates at  
$\nu_{mn}$ and all of its harmonics.
For a high-$Q$ mechanical oscillator, the dynamics are well described by 
\begin{equation}
\ddot{q}_{mn}+(\Gamma_{mn}^{intr}+\Gamma_{mn}^{RP}(\{\chi_{rs}\})) \dot{q}_{mn}+\omega_{mn}^2 q_{mn}=\frac{F_{th}(t)}{m_{eff}}
\label{eqn: modifiedq}
\end{equation}
Here $\omega_{mn}=2\pi\nu_{mn}$, and 
$\Gamma_{mn}^{intr}$ is the intrinsic damping of mode $mn$, related to the
mechanical quality factor $Q_{mn}$ by $\Gamma_{mn}^{intr}=\omega_{mn}/Q_{mn}$. $F_{th}(t)$ is the thermal Langevin force, with spectral density $S_F(\omega)=4k_B Tm_{eff}\Gamma_{mn}^{intr}$.
The  $\Gamma_{mn}^{RP}(\{\chi_{rs}\})$ are optical modifications to the damping of mode $mn$;  
modifications to the $\omega_{mn}$ are also present but not significant here.  In general, $\Gamma_{mn}^{RP}(\{\chi_{rs}\})$ depends on the set of amplitudes $\{\chi_{rs}\}$ of all of the modes.  

If the amplitudes are all small ($\chi_{mn}<<1$), 
only the first-order sidebands need be considered.  In this case $\Gamma_{mn}^{RP}$ can be shown to be independent of the $\{\chi_{rs}\}$, and the optical damping works independently for each mode.  
In previous work we have optically
cooled hundreds of mechanical modes simultaneously\cite{Kemiktarak_NJP_2012}. 


The situation $\Gamma^{RP}_{mn}<0$ corresponds to antidamping, or optically-furnished mechanical gain, 
and is obtained by blue detuning ($\delta\nu>0$).
If the optical gain exceeds the intrinsic damping, $-\Gamma^{RP}_{mn}>\Gamma^{intr}_{mn}$,
the amplitude rings up from its thermal value, and the first-order theory loses validity.
Indeed, as the amplitude of each mode grows, it suppresses the gain of all of the other modes as
determined by the rates $\Gamma^{RP}_{mn}(\{\chi_{rs}\})$. 
This phenomenon of intermode gain suppression has a dramatic signature: it causes an
antidamped system to oscillate on a single mode, even if the unsaturated mechanical gain exceeds the oscillation
threshold for more than one mode.

To see this, we start by calculating $\Gamma^{RP}$ for the case of
two low-order nondegenerate modes, where we use the letters $A$ and $B$ to label mode indices $mn$.  For
$\omega_0<<a$, we find\cite{calculation} 
the gain of mode $A$ to be 
\begin{eqnarray*}
&&\Gamma^{RP}_A(\chi_A,\chi_B)={\cal C}\,P_{in}\left(\frac{\phi_A^2(x_0,y_0)}{\nu_A^2}\right) 
\times \mathrm{Im}\left\{\frac{1}{\chi_A}\sum_{k,l=-\infty}^{\infty}\right.  \\
&&\left.  
\frac{ J_k(\chi_A)J_{k-1}(\chi_A)}{\gamma-i(\delta\omega-(k\omega_A+l\omega_B))}
\frac{J_l^2(\chi_B)}{\gamma+i(\delta\omega-((k-1)\omega_A+l\omega_B))}\right\} 
\end{eqnarray*}
where $\chi_A$ and $\chi_B$ describe the oscillation amplitudes, ${\cal C}=4T_1\Delta\nu^3\,/(\pi m_{eff}\lambda c)$, and the $J_k$ are Bessel functions.  Clearly the gain of mode~$A$ depends on the oscillation amplitude of mode~$B$. The single-mode case, which we consider initially, can be obtained
by taking $\chi_B\rightarrow 0$, $\Gamma^{RP}_A(\chi_A)\equiv\Gamma^{RP}_A(\chi_A,0)$.

\begin{figure}
\includegraphics[width=3.38in,keepaspectratio]{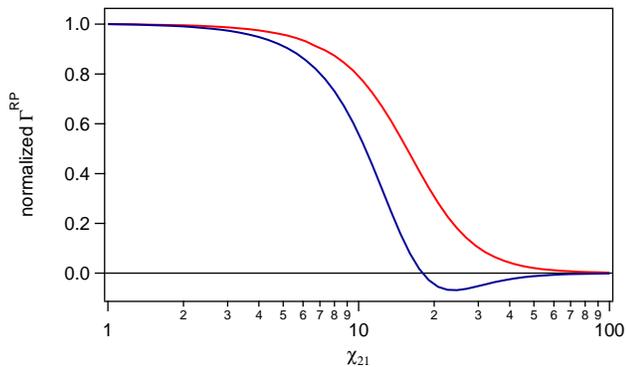}
\caption{Gain saturation vs oscillation amplitude, calculated for detuning $\delta\omega=0.67\gamma$.  {\bf Red:} Single-mode normalized antidamping $\Gamma^{RP}_{21}(\chi_{21})/\Gamma^{RP}_{21}(0)$ of mode (2,1) vs dimensionless amplitude $\chi_{21}$.
{\bf Blue:} Two-mode normalized (anti)damping of mode (1,2) when it is oscillating weakly,
$\Gamma^{RP}_{12}(0,\chi_{21})/\Gamma^{RP}_{12}(0,0)$, for arbitrary $\chi_{21}$.
The gain of the (1,2) mode switches sign for large $\chi_{21}$.}
\label{fig:saturation}
\end{figure}

The single-mode oscillation threshold condition is given by $\Gamma^{RP}_A(0)=-\Gamma^{intr}_A$, and the 
steady-state oscillation amplitude $\chi_A$ is given by $\Gamma^{RP}_A(\chi_A)=-\Gamma^{intr}_A$.
Fig.~\ref{fig:saturation} shows (red curve) the gain of mode $(m,n)=(2,1)$ normalized to its small-amplitude
value, $\Gamma^{RP}_{21}(\chi_{21})/\Gamma^{RP}_{21}(0)$.  We have taken a detuning of $\delta\omega=0.67\gamma$ and a mechanical frequency of $\omega_{21}=0.07\,\gamma$, corresponding to values used in our experiment.
This curve shows gain saturation, as expected: 
$\Gamma^{RP}$ drops to half of the small-amplitude value for an oscillation amplitude of $\chi\approx 16$.

{\em Mode competition. -- }
More interesting phenomena arise when we consider how the gain of mode $(1,2)$ is affected by the amplitude
of mode $(2,1)$.  Fig.~\ref{fig:saturation} (blue curve) shows the gain of the (1,2) mode, 
in the limit of small oscillation amplitude $\chi_{12}$, as a function of $\chi_{21}$.  
The curve exhibits two key features: the gain of the weakly-oscillating mode (1,2) diminishes more rapidly with amplitude $\chi_{21}$ than that of the stronger mode (2,1), and,
when the (2,1) mode oscillates with $\chi_{21}>18$, the gain of the weak mode actually switches sign and provides damping.  Qualitatively similar behavior (not shown) is found for the gain of the (2,1) mode as a function of $\chi_{12}$.

In our experiment, the mode with the lowest threshold power is the (1,2) mode, with $\nu_{12}=192\,{\rm kHz}$,
$\Gamma_{12}^{intr}=2.5(2)\,\mathrm{s}^{-1}$, and geometrical coupling $\phi_{1,2}(x_0,y_0)=0.62(3)$.
The (2,1) mode, with $\nu_{21}=207\,{\rm kHz}$, $\Gamma_{21}^{intr}=4.8(3)\,\mathrm{s}^{-1}$ and $\phi_{2,1}(x_0,y_0)=0.83(2)$, has a slightly higher threshold power, $P^{thresh}_{2,1}=1.05\,P^{thresh}_{1,2}$,
so for incident laser power $P^{thresh}_{1,2}<P_{in}<P^{thresh}_{2,1}$, the (1,2) mode is the only
one that will oscillate.  
The (2,1) mode is, however, better coupled to the radiation pressure, and for 
$P_{in}>>P^{thresh}_{2,1}$, the net small-amplitude gain, 
$-(\Gamma_{21}^{RP}(\{\chi_{rs}^{thermal}\})+\Gamma_{21}^{intr})$, is found to be
1.8 times larger than the corresponding gain for the (1,2) mode. For large pump powers, then, the (2,1) mode  more quickly rings up from thermal amplitude, and as it does, the gain for the more weakly oscillating (1,2) mode is suppressed, as indicated in Fig.~\ref{fig:saturation}.  Thus by appropriate choice of pump power, it is
possible to deterministically force either the (1,2) or (2,1) mode to oscillate.  With different experimental
parameters, we have similarly been able to force the (1,1) and (2,2) modes to oscillate.

\begin{figure}
\includegraphics[width=3.38in,keepaspectratio]{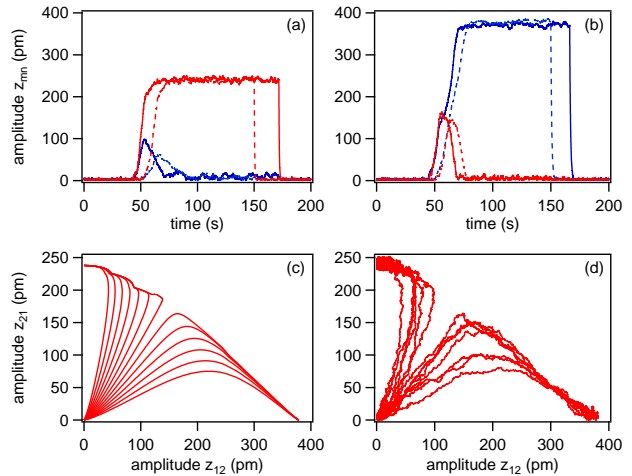}
\caption{(a), (b): Time dependence of mode competition between (1,2) and (2,1) modes, where outcome is
unpredictable.  Blue: (1,2) mode.  Red: (2,1) mode.  Solid lines: experiment; pump 
is turned on at $t\approx 50$~s and turned off at $t\approx 175$~s. Dashed lines: Simulations; pump on/off
times slightly different for clarity.
(c): Set of simulations for slightly different initial conditions; no thermal excitation. (d): Measured set of 13 trajectories such as those shown in (a) and (b).}
\label{fig:competition}
\end{figure}

For pump powers exceeding the oscillation threshold of both modes (1,2) and (2,1), but low enough
that the net small-amplitude gains for the two modes are comparable, it is not possible to predict which mode will oscillate in steady state.  We study the time dependence of the mode competition by sending the signal from the Michelson
interferometer into 
lockin amplifiers 
referenced to $\nu_{12}$ and 
$\nu_{21}$. 
Fig.~\ref{fig:competition}a and~\ref{fig:competition}b (solid lines) show typical  amplitudes $z_{mn}(t)$ for such experiments.  We switch on the pump at time  
$t\approx50\,\mathrm{s}$, with detuning $\delta\omega=0.67\gamma$
and power slightly larger than $P^{thresh}_{2,1}$, and switch it back off at $t\approx175\,\mathrm{s}$. 
The amplitudes of the (1,2)
and (2,1) modes both initially grow, but after several seconds, 
one mode grows until its gain is saturated, while the growth of the other mode is quenched.  Independent
measurements 
confirm that the oscillations of all other modes are likewise quenched.

From the amplitudes of the steady-state oscillation in Fig.~\ref{fig:competition}a and 
Fig.~\ref{fig:competition}b ($75\,\mathrm{s}<t<175\,\mathrm{s}$), we infer $\chi_{21}=7.4$ and $\chi_{12}=9.2$, respectively.  
From the calculated dependence of gains on $\chi_{21}$ (Fig.~\ref{fig:saturation}) and $\chi_{12}$,
one infers $P_{in}=1.18(2)\,P^{thresh}_{1,2}$, and one also finds
the net damping $\Gamma_{mn}^{net}=\Gamma_{mn}^{intr}+\Gamma_{mn}^{RP}(\{\chi_{rs}\})$ of the modes
that are quenched to be $\Gamma_{12}^{net}\approx0.2\,\mathrm{s}^{-1}$ and $\Gamma_{21}^{net}\approx1.1\,\mathrm{s}^{-1}$.
The fact that the net damping of the (1,2) mode is so small manifests itself in the size of the
fluctuations of the quenched (1,2) mode while the (2,1) mode is oscillating (Fig.~\ref{fig:competition}a, 
$75\,\mathrm{s}<t<175\,\mathrm{s}$), that are well above the thermal level ($t<45\,\mathrm{s}$).
The dashed curves in the figures show the results of simulations based on numerical
integration of Eq.~(\ref{eqn: modifiedq}), in which the thermal force~$F_{th}(t)$ is modeled by means
of a memoryless Gaussian stochastic process.
As in the experiment, the mode that ultimately oscillates cannot be predicted in advance. 
The only adjustable parameter in the simulation is the pump laser power, taken to be $P_{in}=1.18\,P_{1,2}^{thresh}$.

Fig.~\ref{fig:competition}c shows a set of trajectories calculated by integrating Eq.~(\ref{eqn: modifiedq})
with $P_{in}=1.18\,P_{1,2}^{thresh}$ for a variety of initial conditions, taking $T=0$
for clarity.  Similar curves were shown 
in the paper by Lamb\cite{lamb_theory_1964}
in his study of multimode operation of an ``optical maser.''  Corresponding curves for thirteen successive realizations of the experiment are shown in Fig.~\ref{fig:competition}d.

\begin{figure}
\includegraphics[width=3.38in,keepaspectratio]{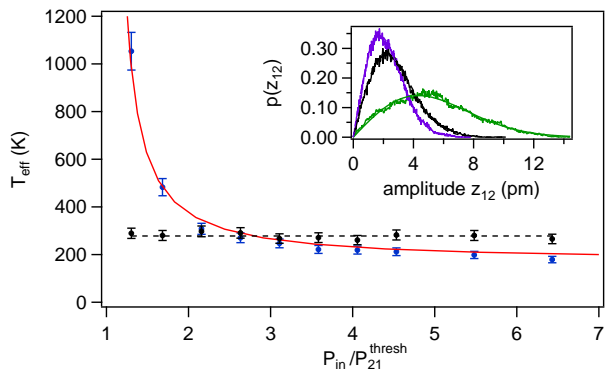}
\caption{Effective temperature of (1,2) mode while (2,1) mode oscillates,
illustrating anomalous cooling for large pump powers.  
Black: Temperature inferred without pump, $T_0=278(21)\,\mathrm{K}$.  Blue circles: Effective temperature with pump.  
Red line: Analytic prediction 
$T_{eff}=293~\mathrm{K}\times \Gamma_{12}^{intr}/(\Gamma_{12}^{intr}+\Gamma_{12}^{RP})$.  {\bf Inset:} Histograms of amplitude of (1,2) mode.  Black: No pump.  Green: Weak pump, $P_{in}=1.3\,P_{2,1}^{thresh}$.  Violet: Strong pump, $P_{in}=6.4\,P_{2,1}^{thresh}$.}
\label{fig:cooling}
\end{figure}

{\em Anomalous cooling. -- }
While the amplitudes of the fluctuations in the quenched mode are above the thermal level
in Fig.~\ref{fig:competition}a and Fig.~\ref{fig:competition}b, the calculated antidamping shown in Fig.~\ref{fig:saturation}
shows that we expect cooling of the quenched mode when the amplitude of the oscillating mode
is large enough.  To study this matter, we set the (2,1) mode into oscillation with a sequence of ten
pump powers from $P_{in}=1.3\,P_{2,1}^{thresh}$ to $P_{in}=6.4\,P_{2,1}^{thresh}$.
For each power, we measured the amplitude $z_{mn}(t)$ of both the (1,2) and (2,1) modes for 450 seconds, then extinguished the pump, allowed the transients to die away, and measured the thermal amplitudes for another 450 seconds.  The inset to Fig.~\ref{fig:cooling} shows histograms of the amplitudes $z_{12}(t)$ for the cases of
no pump, low ($P_{in}=1.3\,P_{2,1}^{thresh}$) and high ($P_{in}=6.4\,P_{2,1}^{thresh}$) pump powers.  In thermal equilibrium, 
the amplitude $z_{mn}$ is distributed according to a Boltzmann distribution
\begin{equation}
p(z_{mn})=\frac{m_{eff}\omega_{mn}^2}{k_B T}z_{mn}\,e^{-\frac{m_{eff}\omega_{mn}^2z_{mn}^2}{2k_B T}}
\label{eqn: Boltzmann}
\end{equation}
Each of the curves in the inset is fit to Eq.~(\ref{eqn: Boltzmann}).  The statistics of the 
set of ten such thermal (pump off) measurements yields a mean of $T_0=278\,\mathrm{K}$ with
a standard deviation of $\approx12\,\mathrm{K}$.  The largest uncertainty in the inferred temperature
arises from $\phi_{12}(x_0,y_0)$, used to infer $z_{12}(t)$, and contributes an uncertainty of 6\%;
adding the statistical contribution in quadrature, we assign an uncertainty of 7.5\% to the 
temperature measurements.  At $P_{in}=1.3\,P_{2,1}^{thresh}$, the statistics 
of the fluctuations in the (1,2) mode, while the (2,1) mode is oscillating, correspond
to an effective temperature of $1040(77)\,\mathrm{K}$.
At $P_{in}=6.4\,P_{2,1}^{thresh}$, the effective temperature is $180(14)\,\mathrm{K}$, illustrating the anomalous cooling predicted in Fig.~\ref{fig:saturation}.  
Fig.~\ref{fig:cooling} shows the effective temperature inferred
from fits to~(\ref{eqn: Boltzmann}) for all ten pump powers. Also shown is the
analytic prediction 
$T_{eff}=293~\mathrm{K}\times \Gamma_{12}^{intr}/(\Gamma_{12}^{intr}+\Gamma_{12}^{RP})$\cite{kippenberg_cavity_2007}.

{\em Conclusion. -- }
We have studied the problem of mode competition in a multimode phonon laser both theoretically and
experimentally.  By using a highly reflective membrane as the end mirror of an optical cavity, in which many mechanical modes are coupled to the intracavity radiation pressure, we demonstrate that oscillation of
one mode tends to ``steal'' gain from more weakly oscillating modes, culminating in single-mode steady-state
operation.  Remarkably, strong oscillation of one mode even causes optical damping of the other modes.
In addition to more fully illuminating the analogy between phonon lasers and their optical counterparts,
the insights gained here can be used to force a particular mode to oscillate when multiple modes are
capable of oscillation, which may be useful as applications of phonon lasers appear.  

\begin{acknowledgments}
We acknowledge useful discussions with Jake Taylor and National Science Foundation support through 
the Physics Frontier Center at the Joint Quantum Institute.  Research performed in part at the NIST 
Center for Nanoscale Science and Technology.
\end{acknowledgments}


\begin{thebibliography}{22}%
\makeatletter
\providecommand \@ifxundefined [1]{%
 \@ifx{#1\undefined}
}%
\providecommand \@ifnum [1]{%
 \ifnum #1\expandafter \@firstoftwo
 \else \expandafter \@secondoftwo
 \fi
}%
\providecommand \@ifx [1]{%
 \ifx #1\expandafter \@firstoftwo
 \else \expandafter \@secondoftwo
 \fi
}%
\providecommand \natexlab [1]{#1}%
\providecommand \enquote  [1]{``#1''}%
\providecommand \bibnamefont  [1]{#1}%
\providecommand \bibfnamefont [1]{#1}%
\providecommand \citenamefont [1]{#1}%
\providecommand \href@noop [0]{\@secondoftwo}%
\providecommand \href [0]{\begingroup \@sanitize@url \@href}%
\providecommand \@href[1]{\@@startlink{#1}\@@href}%
\providecommand \@@href[1]{\endgroup#1\@@endlink}%
\providecommand \@sanitize@url [0]{\catcode `\\12\catcode `\$12\catcode
  `\&12\catcode `\#12\catcode `\^12\catcode `\_12\catcode `\%12\relax}%
\providecommand \@@startlink[1]{}%
\providecommand \@@endlink[0]{}%
\providecommand \url  [0]{\begingroup\@sanitize@url \@url }%
\providecommand \@url [1]{\endgroup\@href {#1}{\urlprefix }}%
\providecommand \urlprefix  [0]{URL }%
\providecommand \Eprint [0]{\href }%
\providecommand \doibase [0]{http://dx.doi.org/}%
\providecommand \selectlanguage [0]{\@gobble}%
\providecommand \bibinfo  [0]{\@secondoftwo}%
\providecommand \bibfield  [0]{\@secondoftwo}%
\providecommand \translation [1]{[#1]}%
\providecommand \BibitemOpen [0]{}%
\providecommand \bibitemStop [0]{}%
\providecommand \bibitemNoStop [0]{.\EOS\space}%
\providecommand \EOS [0]{\spacefactor3000\relax}%
\providecommand \BibitemShut  [1]{\csname bibitem#1\endcsname}%
\let\auto@bib@innerbib\@empty
\bibitem [{\citenamefont {Tucker}(1961)}]{Tucker_Ruby}%
  \BibitemOpen
  \bibfield  {author} {\bibinfo {author} {\bibfnamefont {E.~B.}\ \bibnamefont
  {Tucker}},\ }\href@noop {} {\bibfield  {journal} {\bibinfo  {journal}
  {Physical Review Letters}\ }\textbf {\bibinfo {volume} {6}} (\bibinfo {year}
  {1961})}\BibitemShut {NoStop}%
\bibitem [{\citenamefont {Ganapolskii}\ and\ \citenamefont
  {Makovetskii}(1974)}]{ganapolskii_resonant_1974}%
  \BibitemOpen
  \bibfield  {author} {\bibinfo {author} {\bibfnamefont {E.~M.}\ \bibnamefont
  {Ganapolskii}}\ and\ \bibinfo {author} {\bibfnamefont {D.~N.}\ \bibnamefont
  {Makovetskii}},\ }\href@noop {} {\bibfield  {journal} {\bibinfo  {journal}
  {Solid state communications}\ }\textbf {\bibinfo {volume} {15}},\ \bibinfo
  {pages} {1249 } (\bibinfo {year} {1974})}\BibitemShut {NoStop}%
\bibitem [{\citenamefont {Ganapolskii}\ and\ \citenamefont
  {Makovetskii}(1994)}]{ganapolskii_resonant_1994}%
  \BibitemOpen
  \bibfield  {author} {\bibinfo {author} {\bibfnamefont {E.~M.}\ \bibnamefont
  {Ganapolskii}}\ and\ \bibinfo {author} {\bibfnamefont {D.~N.}\ \bibnamefont
  {Makovetskii}},\ }\href
  {http://www.sciencedirect.com/science/article/pii/003810989490054X}
  {\bibfield  {journal} {\bibinfo  {journal} {Solid state communications}\
  }\textbf {\bibinfo {volume} {90}},\ \bibinfo {pages} {501 } (\bibinfo {year}
  {1994})}\BibitemShut {NoStop}%
\bibitem [{\citenamefont {Tilstra}\ \emph {et~al.}(2003)\citenamefont
  {Tilstra}, \citenamefont {Arts},\ and\ \citenamefont
  {de~Wijn}}]{tilstra_coherence_2003}%
  \BibitemOpen
  \bibfield  {author} {\bibinfo {author} {\bibfnamefont {L.}~\bibnamefont
  {Tilstra}}, \bibinfo {author} {\bibfnamefont {A.}~\bibnamefont {Arts}}, \
  and\ \bibinfo {author} {\bibfnamefont {H.}~\bibnamefont {de~Wijn}},\ }\href
  {\doibase 10.1103/PhysRevB.68.144302} {\bibfield  {journal} {\bibinfo
  {journal} {Physical Review B}\ }\textbf {\bibinfo {volume} {68}},\ \bibinfo {pages} {144302 } (\bibinfo  {year} {2003})}\BibitemShut {NoStop}%
\bibitem [{\citenamefont {Tilstra}\ \emph {et~al.}(2007)\citenamefont
  {Tilstra}, \citenamefont {Arts},\ and\ \citenamefont
  {de~Wijn}}]{tilstra_optically_2007}%
  \BibitemOpen
  \bibfield  {author} {\bibinfo {author} {\bibfnamefont {L.}~\bibnamefont
  {Tilstra}}, \bibinfo {author} {\bibfnamefont {A.}~\bibnamefont {Arts}}, \
  and\ \bibinfo {author} {\bibfnamefont {H.}~\bibnamefont {de~Wijn}},\ }\href
  {\doibase 10.1103/PhysRevB.76.024302} {\bibfield  {journal} {\bibinfo
  {journal} {Physical Review B}\ }\textbf {\bibinfo {volume} {76}},\ \bibinfo {pages} {024302 }  (\bibinfo  {year} {2007})}\BibitemShut {NoStop}%
\bibitem [{\citenamefont {Vahala}\ \emph {et~al.}(2009)\citenamefont {Vahala},
  \citenamefont {Herrmann}, \citenamefont {Kn\"{u}nz}, \citenamefont {Batteiger},
  \citenamefont {Saathoff}, \citenamefont {H\"{a}nsch},\ and\ \citenamefont
  {Udem}}]{vahala_phonon_2009}%
  \BibitemOpen
  \bibfield  {author} {\bibinfo {author} {\bibfnamefont {K.}~\bibnamefont
  {Vahala}}, \bibinfo {author} {\bibfnamefont {M.}~\bibnamefont {Herrmann}},
  \bibinfo {author} {\bibfnamefont {S.}~\bibnamefont {Kn\"{u}nz}}, \bibinfo
  {author} {\bibfnamefont {V.}~\bibnamefont {Batteiger}}, \bibinfo {author}
  {\bibfnamefont {G.}~\bibnamefont {Saathoff}}, \bibinfo {author}
  {\bibfnamefont {T.~W.}\ \bibnamefont {H\"{a}nsch}}, \ and\ \bibinfo {author}
  {\bibfnamefont {T.}~\bibnamefont {Udem}},\ }\href {\doibase
  10.1038/nphys1367} {\bibfield  {journal} {\bibinfo  {journal} {Nature
  Physics}\ }\textbf {\bibinfo {volume} {5}},\ \bibinfo {pages} {682} (\bibinfo
  {year} {2009})}\BibitemShut {NoStop}%
\bibitem [{\citenamefont {Khurgin}\ \emph {et~al.}(2012)\citenamefont
  {Khurgin}, \citenamefont {Pruessner}, \citenamefont {Stievater},\ and\
  \citenamefont {Rabinovich}}]{khurgin_laser-rate-equation_2012}%
  \BibitemOpen
  \bibfield  {author} {\bibinfo {author} {\bibfnamefont {J.~B.}\ \bibnamefont
  {Khurgin}}, \bibinfo {author} {\bibfnamefont {M.~W.}\ \bibnamefont
  {Pruessner}}, \bibinfo {author} {\bibfnamefont {T.~H.}\ \bibnamefont
  {Stievater}}, \ and\ \bibinfo {author} {\bibfnamefont {W.~S.}\ \bibnamefont
  {Rabinovich}},\ }\href {\doibase 10.1103/PhysRevLett.108.223904} {\bibfield
  {journal} {\bibinfo  {journal} {Physical Review Letters}\ }\textbf {\bibinfo
  {volume} {108}},\ \bibinfo {pages} {223904 }  (\bibinfo {year} {2012})}\BibitemShut {NoStop}%
\bibitem [{\citenamefont {Bagheri}\ \emph {et~al.}(2011)\citenamefont
  {Bagheri}, \citenamefont {Poot}, \citenamefont {Li}, \citenamefont
  {Pernice},\ and\ \citenamefont {Tang}}]{bagheri_dynamic_2011}%
  \BibitemOpen
  \bibfield  {author} {\bibinfo {author} {\bibfnamefont {M.}~\bibnamefont
  {Bagheri}}, \bibinfo {author} {\bibfnamefont {M.}~\bibnamefont {Poot}},
  \bibinfo {author} {\bibfnamefont {M.}~\bibnamefont {Li}}, \bibinfo {author}
  {\bibfnamefont {W.~P.~H.}\ \bibnamefont {Pernice}}, \ and\ \bibinfo {author}
  {\bibfnamefont {H.~X.}\ \bibnamefont {Tang}},\ }\href {\doibase
  10.1038/nnano.2011.180} {\bibfield  {journal} {\bibinfo  {journal} {Nature
  Nanotechnology}\ }\textbf {\bibinfo {volume} {6}},\ \bibinfo {pages} {726}
  (\bibinfo {year} {2011})}\BibitemShut {NoStop}%
\bibitem [{\citenamefont {Metzger}\ \emph {et~al.}(2008)\citenamefont
  {Metzger}, \citenamefont {Ludwig}, \citenamefont {Neuenhahn}, \citenamefont
  {Ortlieb}, \citenamefont {Favero}, \citenamefont {Karrai},\ and\
  \citenamefont {Marquardt}}]{metzger_self-induced_2008}%
  \BibitemOpen
  \bibfield  {author} {\bibinfo {author} {\bibfnamefont {C.}~\bibnamefont
  {Metzger}}, \bibinfo {author} {\bibfnamefont {M.}~\bibnamefont {Ludwig}},
  \bibinfo {author} {\bibfnamefont {C.}~\bibnamefont {Neuenhahn}}, \bibinfo
  {author} {\bibfnamefont {A.}~\bibnamefont {Ortlieb}}, \bibinfo {author}
  {\bibfnamefont {I.}~\bibnamefont {Favero}}, \bibinfo {author} {\bibfnamefont
  {K.}~\bibnamefont {Karrai}}, \ and\ \bibinfo {author} {\bibfnamefont
  {F.}~\bibnamefont {Marquardt}},\ }\href {\doibase
  10.1103/PhysRevLett.101.133903} {\bibfield  {journal} {\bibinfo  {journal}
  {Physical Review Letters}\ }\textbf {\bibinfo {volume} {101}},\ \bibinfo {pages} {133903 } (\bibinfo  {year} {2008})}\BibitemShut {NoStop}%
\bibitem [{\citenamefont {Kippenberg}\ and\ \citenamefont
  {Vahala}(2007)}]{kippenberg_cavity_2007}%
  \BibitemOpen
  \bibfield  {author} {\bibinfo {author} {\bibfnamefont {T.~J.}\ \bibnamefont
  {Kippenberg}}\ and\ \bibinfo {author} {\bibfnamefont {K.~J.}\ \bibnamefont
  {Vahala}},\ }\href {\doibase 10.1364/OE.15.017172} {\bibfield  {journal}
  {\bibinfo  {journal} {Optics Express}\ }\textbf {\bibinfo {volume} {15}},\
  \bibinfo {pages} {17172} (\bibinfo {year} {2007})}\BibitemShut {NoStop}%
\bibitem [{\citenamefont {Grudinin}\ \emph {et~al.}(2010)\citenamefont
  {Grudinin}, \citenamefont {Lee}, \citenamefont {Painter},\ and\ \citenamefont
  {Vahala}}]{grudinin_phonon_2010}%
  \BibitemOpen
  \bibfield  {author} {\bibinfo {author} {\bibfnamefont {I.~S.}\ \bibnamefont
  {Grudinin}}, \bibinfo {author} {\bibfnamefont {H.}~\bibnamefont {Lee}},
  \bibinfo {author} {\bibfnamefont {O.}~\bibnamefont {Painter}}, \ and\
  \bibinfo {author} {\bibfnamefont {K.~J.}\ \bibnamefont {Vahala}},\ }\href
  {\doibase 10.1103/PhysRevLett.104.083901} {\bibfield  {journal} {\bibinfo
  {journal} {Physical Review Letters}\ }\textbf {\bibinfo {volume} {104}},\ \bibinfo {pages} {083901 } (\bibinfo {year} {2010})}\BibitemShut{NoStop}%
\bibitem [{\citenamefont {Okamoto}\ \emph {et~al.}(2011)\citenamefont
  {Okamoto}, \citenamefont {Ito}, \citenamefont {Onomitsu}, \citenamefont
  {Sanada}, \citenamefont {Gotoh}, \citenamefont {Sogawa},\ and\ \citenamefont
  {Yamaguchi}}]{okamoto_vibration_2011}%
  \BibitemOpen
  \bibfield  {author} {\bibinfo {author} {\bibfnamefont {H.}~\bibnamefont
  {Okamoto}}, \bibinfo {author} {\bibfnamefont {D.}~\bibnamefont {Ito}},
  \bibinfo {author} {\bibfnamefont {K.}~\bibnamefont {Onomitsu}}, \bibinfo
  {author} {\bibfnamefont {H.}~\bibnamefont {Sanada}}, \bibinfo {author}
  {\bibfnamefont {H.}~\bibnamefont {Gotoh}}, \bibinfo {author} {\bibfnamefont
  {T.}~\bibnamefont {Sogawa}}, \ and\ \bibinfo {author} {\bibfnamefont
  {H.}~\bibnamefont {Yamaguchi}},\ }\href {\doibase
  10.1103/PhysRevLett.106.036801} {\bibfield  {journal} {\bibinfo  {journal}
  {Physical Review Letters}\ }\textbf {\bibinfo {volume} {106}},\ \bibinfo {pages} {036801 } (\bibinfo  {year} {2011})}\BibitemShut {NoStop}%
\bibitem [{\citenamefont {Bargatin}\ and\ \citenamefont
  {Roukes}(2003)}]{bargatin_nanomechanical_2003}%
  \BibitemOpen
  \bibfield  {author} {\bibinfo {author} {\bibfnamefont {I.}~\bibnamefont
  {Bargatin}}\ and\ \bibinfo {author} {\bibfnamefont {M.}~\bibnamefont
  {Roukes}},\ }\href {\doibase 10.1103/PhysRevLett.91.138302} {\bibfield
  {journal} {\bibinfo  {journal} {Physical Review Letters}\ }\textbf {\bibinfo
  {volume} {91}},\ \bibinfo {pages} {138302 } (\bibinfo {year} {2003})}\BibitemShut {NoStop}%
\bibitem [{\citenamefont {Bennett}\ and\ \citenamefont
  {Clerk}(2006)}]{Bennett2006}%
  \BibitemOpen
  \bibfield  {author} {\bibinfo {author} {\bibfnamefont {S.~D.}\ \bibnamefont
  {Bennett}}\ and\ \bibinfo {author} {\bibfnamefont {A.~A.}\ \bibnamefont
  {Clerk}},\ }\href {\doibase 10.1103/PhysRevB.74.201301} {\bibfield  {journal}
  {\bibinfo  {journal} {Phys. Rev. B}\ }\textbf {\bibinfo {volume} {74}},\
  \bibinfo {pages} {201301} (\bibinfo {year} {2006})}\BibitemShut {NoStop}%
\bibitem [{\citenamefont {Mahboob}\ \emph {et~al.}(2013)\citenamefont
  {Mahboob}, \citenamefont {Nishiguchi}, \citenamefont {Fujiwara},\ and\
  \citenamefont {Yamaguchi}}]{mahboob_phonon_2013}%
  \BibitemOpen
  \bibfield  {author} {\bibinfo {author} {\bibfnamefont {I.}~\bibnamefont
  {Mahboob}}, \bibinfo {author} {\bibfnamefont {K.}~\bibnamefont {Nishiguchi}},
  \bibinfo {author} {\bibfnamefont {A.}~\bibnamefont {Fujiwara}}, \ and\
  \bibinfo {author} {\bibfnamefont {H.}~\bibnamefont {Yamaguchi}},\ }\href
  {\doibase 10.1103/PhysRevLett.110.127202} {\bibfield  {journal} {\bibinfo
  {journal} {Physical Review Letters}\ }\textbf {\bibinfo {volume} {110}},\
  \bibinfo {pages} {127202} (\bibinfo {year} {2013})}\BibitemShut {NoStop}%
\bibitem [{\citenamefont {Kn\"{u}nz}\ \emph {et~al.}(2010)\citenamefont
  {Kn\"{u}nz}, \citenamefont {Herrmann}, \citenamefont {Batteiger},
  \citenamefont {Saathoff}, \citenamefont {H\"{a}nsch}, \citenamefont
  {Vahala},\ and\ \citenamefont {Udem}}]{knuenz_injection_2010}%
  \BibitemOpen
  \bibfield  {author} {\bibinfo {author} {\bibfnamefont {S.}~\bibnamefont
  {Kn\"{u}nz}}, \bibinfo {author} {\bibfnamefont {M.}~\bibnamefont {Herrmann}},
  \bibinfo {author} {\bibfnamefont {V.}~\bibnamefont {Batteiger}}, \bibinfo
  {author} {\bibfnamefont {G.}~\bibnamefont {Saathoff}}, \bibinfo {author}
  {\bibfnamefont {T.}~\bibnamefont {H\"{a}nsch}}, \bibinfo {author}
  {\bibfnamefont {K.}~\bibnamefont {Vahala}}, \ and\ \bibinfo {author}
  {\bibfnamefont {T.}~\bibnamefont {Udem}},\ }\href {\doibase
  10.1103/PhysRevLett.105.013004} {\bibfield  {journal} {\bibinfo  {journal}
  {Physical Review Letters}\ }\textbf {\bibinfo {volume} {105}},\ \bibinfo {pages} {013004 } (\bibinfo  {year} {2010})}\BibitemShut {NoStop}%
\bibitem [{\citenamefont {Lamb}(1964)}]{lamb_theory_1964}%
  \BibitemOpen
  \bibfield  {author} {\bibinfo {author} {\bibfnamefont {W.~E.}\ \bibnamefont
  {Lamb}},\ }\href {\doibase 10.1103/PhysRev.134.A1429} {\bibfield  {journal}
  {\bibinfo  {journal} {Physical Review}\ }\textbf {\bibinfo {volume} {134}},\
  \bibinfo {pages} {A1429} (\bibinfo {year} {1964})}\BibitemShut {NoStop}%
\bibitem [{\citenamefont {Siegman}(1986)}]{Siegman}%
  \BibitemOpen
  \bibfield  {author} {\bibinfo {author} {\bibfnamefont {A.}~\bibnamefont
  {Siegman}},\ }\enquote {\bibinfo {title} {Lasers},}\ \ (\bibinfo  {publisher}
  {University Science Books, Sausalito},\ \bibinfo {year} {1986})\BibitemShut
  {NoStop}%
\bibitem [{\citenamefont {Kemiktarak}\ \emph
  {et~al.}(2012{\natexlab{a}})\citenamefont {Kemiktarak}, \citenamefont
  {Metcalfe}, \citenamefont {Durand},\ and\ \citenamefont
  {Lawall}}]{kemiktarak_mechanically_2012}%
  \BibitemOpen
  \bibfield  {author} {\bibinfo {author} {\bibfnamefont {U.}~\bibnamefont
  {Kemiktarak}}, \bibinfo {author} {\bibfnamefont {M.}~\bibnamefont
  {Metcalfe}}, \bibinfo {author} {\bibfnamefont {M.}~\bibnamefont {Durand}}, \
  and\ \bibinfo {author} {\bibfnamefont {J.}~\bibnamefont {Lawall}},\ }\href
  {\doibase 10.1063/1.3684248} {\bibfield  {journal} {\bibinfo  {journal}
  {Applied Physics Letters}\ }\textbf {\bibinfo {volume} {100}},\ \bibinfo
  {pages} {061124} (\bibinfo {year} {2012}{\natexlab{a}})}\BibitemShut
  {NoStop}%
\bibitem [{\citenamefont {Kemiktarak}\ \emph
  {et~al.}(2012{\natexlab{b}})\citenamefont {Kemiktarak}, \citenamefont
  {Durand}, \citenamefont {Metcalfe},\ and\ \citenamefont
  {Lawall}}]{Kemiktarak_NJP_2012}%
  \BibitemOpen
  \bibfield  {author} {\bibinfo {author} {\bibfnamefont {U.}~\bibnamefont
  {Kemiktarak}}, \bibinfo {author} {\bibfnamefont {M.}~\bibnamefont {Durand}},
  \bibinfo {author} {\bibfnamefont {M.}~\bibnamefont {Metcalfe}}, \ and\
  \bibinfo {author} {\bibfnamefont {J.}~\bibnamefont {Lawall}},\ }\href@noop {}
  {\bibfield  {journal} {\bibinfo  {journal} {New Journal of Physics}\ }\textbf
  {\bibinfo {volume} {14}},\ \bibinfo {pages} {125010} (\bibinfo {year}
  {2012}{\natexlab{b}})}\BibitemShut {NoStop}%
\bibitem [{\citenamefont {Timoshenko}(1937)}]{Timoshenko}%
  \BibitemOpen
  \bibfield  {author} {\bibinfo {author} {\bibnamefont {Timoshenko}},\
  }\href@noop {} {\emph {\bibinfo {title} {Vibration Problems in
  Engineering}}}\ (\bibinfo  {publisher} {D. Van Nostrand Company, New York},\ \bibinfo
  {year} {1937})\BibitemShut {NoStop}%
\bibitem [{cal()}]{calculation}%
  \BibitemOpen
  \href@noop {} {}\bibinfo {note} {See Supplemental Material for details of
  calculation}\BibitemShut {NoStop}%
\end{thebibliography}

%

\end{document}


\maketitle
\begin{center}
$^1$Joint Quantum Institute, University of Maryland, College Park, MD 20742, USA \\
$^2$National Institute of Standards and Technology, 100 Bureau Drive, Gaithersburg, MD 20899, USA
\end{center}

\section{Generalized force from radiation pressure in Gaussian beam}
Newton's equation of motion for $q_{mn}$ is
\begin{equation}
\ddot{q}_{mn}+\Gamma_{mn}^{intr} \dot{q}_{mn}+\omega_{mn}^2 q_{mn}=\frac{F_{mn}(t)}{m_{eff}}+\frac{F_{th}(t)}{m_{eff}}
\label{eqn: q}
\end{equation}
where each mode is driven by the generalized force
\begin{equation}
F_{mn}(t)=\int\int f(x,y,t)\phi_{mn}(x,y)\,dx\,dy
\label{eqn: Fmn}
\end{equation}
We take the mode functions $\phi_{mn}(x,y)$ to be those of a uniform square membrane,
$\phi_{mn}(x,y)=\sin\frac{m\pi x}{a}\sin\frac{n\pi y}{a}$, and the
radiation pressure 
force per unit area $f(x,y,t)$ to have a Gaussian spatial distribution, with spot size
$\omega_0$, centered at $(x_0,y_0)$:
\begin{equation}
f(x,y,t)=\frac{2F^{RP}(t)}{\pi\omega_0^2}e^{-2((x-x_0)^2+(y-y_0)^2)/\omega_0^2}
\end{equation}
Here $F^{RP}(t)$ is the total radiation pressure force
\begin{equation}
F^{RP}(t)=\int\int f(x,y,t)\,dx\,dy
\end{equation}
equal to
\begin{equation}
F^{RP}(t)=\frac{2}{c}|u(t)|^2 
\end{equation}
where $u(t)$ is the amplitude of the  electric field circulating in the cavity,
and units are chosen such that $|u(t)|^2$ has dimensions of power.
Evaluating the integral in Eqn.~(\ref{eqn: Fmn}),
\begin{eqnarray}
F_{mn}(t)&=&F^{RP}(t) \exp\left(\frac{-\pi^2\omega_0^2}{8a^2}\left(m^2+n^2\right)\right)u_{mn}(x_0,y_0) \\
&\approx&F^{RP}(t) \phi_{mn}(x_0,y_0) \label{eqn: genforce}
\end{eqnarray}
where the final approximation assumes low-order modes $m$ and $n$ and $\omega_0<<a$.

\section{Dynamics of circulating optical field}
Substituting our normal mode decomposition of the membrane motion
\begin{equation}
z(t)=\sum_{m,n=1}^{\infty}q_{mn}(t)\phi_{mn}(x_0,y_0)
\end{equation}
into the equation for the amplitude $u(t)$ of the field circulating in the Fabry-Perot cavity,
and making the ansatz $q_{mn}(t)=z_{mn}\sin \omega_{mn} t$, we obtain
\begin{equation}
\dot{u}(t)+\left(\gamma - i\,\left(\delta\omega+\frac{2 \Delta\nu}{\lambda}\sum_{m,n=1}^{\infty}
\frac{\omega_{mn}}{\nu_{mn}}z_{mn}\,\phi_{mn}(x_0,y_0)\sin \omega_{mn} t \right)\right)\,u(t)=i\Delta\nu\,\sqrt{T_1\,P_{in}} 
\end{equation}
Note that $z_{mn}$ is the amplitude of the membrane oscillation at antinodes of the mechanical 
mode $mn$, and $z_{mn}\,\phi_{mn}(x_0,y_0)$ is the amplitude at the center of the optical spot.
Taking $\alpha=\gamma - i\,\delta\omega$ and defining the dimensionless parameter
\begin{equation}
\chi_{mn}\equiv 2\,\frac{ z_{mn}\phi_{mn}(x_0,y_0)}{\lambda}\frac{\Delta\nu}{\nu_{mn}}
\end{equation}
where $\omega_{mn}=2\pi\nu_{mn}$, the equation of evolution for $u$ becomes
\begin{equation}
\dot{u}(t)+\left(\alpha- i\sum_{m,n=1}^{\infty}\omega_{mn}\chi_{mn}\sin \omega_{mn} t \right)\,u(t)=i\Delta\nu\,\sqrt{T_1\,P_{in}} 
\label{eqn: field evolution}
\end{equation}
We consider only two modes, which we label $A$ and $B$, so that Eq.~(\ref{eqn: field evolution}) becomes
\begin{equation}
\dot{u}+\left(\alpha - i\,(\omega_{A}\chi_{A}\sin\omega_{A} t+\omega_{B}\chi_{B}\sin\omega_{B} t)\right)\,u=i\Delta\nu\,\sqrt{T_1\,P_{in}} 
\label{eqn: field evolution 2}
\end{equation}
Using an integrating factor, the steady-state solution of this equation is
\begin{eqnarray}
u(t)&=&i\sqrt{T_1\,P_{in}}\, \Delta\nu\,e^{-\alpha t-i(\chi_{A} \cos\omega_{A} t+\chi_{B} \cos\omega_{B} t)}\,\int^t 
e^{\alpha t'+i(\chi_{A} \cos\omega_{A} t'+\chi_{B} \cos\omega_{B} t')}\,dt' \\
&=&i\sqrt{T_1\,P_{in}}\,\Delta\nu\,e^{-i(\chi_{A} \cos\omega_{A} t+\chi_{B} \cos\omega_{B} t)}\,
\sum_{k,l=-\infty}^{\infty}i^{k+l} J_k(\chi_{A})J_l(\chi_{B})\frac{e^{i(k\omega_{A}+l\omega_{B}) t}}
{\alpha+i(k\omega_{A}+l\omega_{B})}
\end{eqnarray}
where we have expanded the exponential in the integrand and used the identity
\begin{equation}
e^{i\chi \cos\phi}=\sum_{k=-\infty}^{\infty}i^k J_k(\chi)e^{ik\phi}
\end{equation}
\section{Radiation pressure force}
The circulating optical power is given by
\[
|u(t)|^2=T_1 P_{in}\,\Delta\nu^2\,\sum_{k,l,s,q=-\infty}^{\infty}i^{s+q}
J_k(\chi_{A})J_l(\chi_{B})J_{k-s}(\chi_{A})J_{l-q}(\chi_{B})
\frac{e^{is\omega_{A} t}}{\alpha+i(k\omega_{A}+l\omega_{B})}
 \frac{e^{iq\omega_{B} t}}{\alpha^*-i((k-s)\omega_{A}+(l-q)\omega_{B})} 
\]
Clearly the radiation pressure force $F^{RP}=\frac{2}{c}|u(t)|^2$ oscillates at all multiples of $\omega_{A}$,
$\omega_{B}$, and an infinite number of sum and difference frequencies.  To determine the (anti)damping
of mode $A$, the terms at $\pm\omega_{A}$ are of particular importance, since they are the only ones that contribute to the cycle-averaged work done on that mode of the mechanical oscillator.  $F^{RP}$ can then be written
\begin{eqnarray*}
F^{RP}(t)&\rightarrow&-\frac{4}{c}T_1\Delta\nu^2\,P_{in}\, \left\{Re\sum_{k,l=-\infty}^{\infty}  
\frac{ J_k(\chi_{A})J_{k-1}(\chi_{A})}{\alpha+i(k\omega_{A}+l\omega_{B})}
\frac{J_l^2(\chi_{B})}{\alpha^*-i((k-1)\omega_{A}+l\omega_{B})}\sin\omega_{A} t \right.  \\
&&\left.+Im\sum_{k,l=-\infty}^{\infty}  
\frac{ J_k(\chi_{A})J_{k-1}(\chi_{A})}{\alpha+i(k\omega_{A}+l\omega_{B})}
\frac{J_l^2(\chi_{B})}{\alpha^*-i((k-1)\omega_{A}+l\omega_{B})}\cos\omega_{A} t 
\right\} +F_{nonresonant}
\end{eqnarray*}
We now use
\begin{eqnarray*}
q_{A}(t)&=&z_{A}\sin \omega_{A} t \\
\dot{q}_{A}(t)&=&\omega_{A} z_{A}\cos\omega_{A} t
\end{eqnarray*}
and Eq.~(\ref{eqn: genforce}) to express the generalized force on mode $A$ as
\begin{eqnarray*}
F_{A}(t)&=&-\frac{4\phi_A(x_0,y_0)}{cz_A}T_1\Delta\nu^2\,P_{in}\, \left\{q_A(t) Re\sum_{k,l=-\infty}^{\infty}  
\frac{ J_k(\chi_{A})J_{k-1}(\chi_{A})}{\alpha+i(k\omega_{A}+l\omega_{B})}
\frac{J_l^2(\chi_{B})}{\alpha^*-i((k-1)\omega_{A}+l\omega_{B})} \right.  \\
&&\left.+\frac{\dot{q}_A(t)}{\omega_{A}}Im\sum_{k,l=-\infty}^{\infty}  
\frac{ J_k(\chi_{A})J_{k-1}(\chi_{A})}{\alpha+i(k\omega_{A}+l\omega_{B})}
\frac{J_l^2(\chi_{B})}{\alpha^*-i((k-1)\omega_{A}+l\omega_{B})}\right\} +F_{nonresonant}
\end{eqnarray*}
Substituting into Eq.~(\ref{eqn: q}) and dropping the nonresonant terms, we find
\begin{equation}
\ddot{q}_{A}+\left(\Gamma_{A}^{intr}+\Gamma_{A}^{RP}(\chi_{A},\chi_{B})\right) \dot{q}_{A}
+\omega_{A}^2 \,q_{A}=\frac{F_{th}(t)}{m_{eff}}
\end{equation}
where an optically-induced modification to the oscillation frequency $\omega_A$ (not relevant
to this work) has been suppressed,
and the optical contribution to the mechanical (anti)damping is given by
\begin{equation}
\Gamma^{RP}_A(\chi_{A},\chi_{B})={\cal C}\,P_{in}\left(\frac{\phi_A^2(x_0,y_0)}{\nu_A^2}\right)\,
Im\left\{\frac{1}{\chi_{A}}\sum_{k,l=-\infty}^{\infty}  
\frac{ J_k(\chi_{A})J_{k-1}(\chi_{A})}{\gamma - i(\delta\omega-(k\omega_{A}+l\omega_{B}))}
\frac{J_l^2(\chi_{B})}{\gamma + i(\delta\omega-((k-1)\omega_{A}+l\omega_{B}))}\right\} 
\label{eqn: GammaRP} 
\end{equation}
with
\[
{\cal C}=\frac{4T_1\Delta\nu^3}{\pi m_{eff}\lambda c}
\]
The single-mode case is obtained by taking $\chi_{B}\rightarrow 0$, and gives
\begin{equation}
\Gamma^{RP}_A(\chi_{A})={\cal C}\,P_{in}\left(\frac{\phi_A^2(x_0,y_0)}{\nu_A^2}\right)\,
Im\left\{\frac{1}{\chi_{A}}\sum_{k=-\infty}^{\infty}  
\frac{ J_k(\chi_{A})}{\gamma - i(\delta\omega-k\omega_{A})}
\frac{J_{k-1}(\chi_{A})}{\gamma + i(\delta\omega-(k-1)\omega_{A})}\right\} \label{eqn: 1modeGammaRP} 
\end{equation}